# Low-loss high-speed speckle reduction using a colloidal dispersion

Brandon Redding[1], Graham Allen[1], Eric R. Dufresne[2], and Hui Cao[1,*]

[1]Department of Applied Physics, Yale University, New Haven, CT 06520
[2]Departments of Mechanical Engineering and Materials Science, Chemical and Environmental Engineering, Physics, and Cell Biology, Yale University, New Haven, CT 06520
*Corresponding author: hui.cao@yale.edu

We present a simple and robust approach to reduce laser speckle, which has limited the adoption of lasers in imaging and display applications. We use colloidal solutions that can quickly reduce speckle contrast due to the Brownian motion of the scattering particles. The high insertion loss associated with propagation through a colloidal solution was overcome by using white paint to cover the sides of the cuvette and an optical fiber to deliver the laser light deep into the colloidal solution, enabling transmission greater than 90%. The diffused laser output followed a Lambertian distribution and produced speckle contrast below 4% at an integration time of 129 μs. The ability for colloidal solutions to achieve fast speckle reduction without power consumption while maintaining high transmission, low cost, a compact size, and a long lifetime, makes our approach useful for a wide range of laser imaging and projection applications. © 2012 Optical Society of America

## 1. Introduction

Lasers offer a number of potential advantages in imaging and display applications [1]. Compared with thermal sources or light emitting diodes (LEDs), lasers offer more power per mode for full-field imaging, enabling higher signal-to-noise ratio images of dynamic phenomena, and wavelength-selective excitations for fluorescent and photochemical imaging. However, the adoption of lasers in imaging applications has been limited due to the presence of coherent artifacts such as laser speckle. Speckle is a consequence of the high spatial coherence intrinsic to most lasers [2,3] and occurs due to interference of light with random phase delays. In imaging applications, scattering caused by surface roughness or environmental index variation introduces these phase delays, resulting in artificial intensity modulations which corrupt image formation and impair human ability to interpret images [4,5]. As a result, there has been considerable effort in developing efficient methods to suppress laser speckle [1,3,6].

While it is possible to generate laser emission with low spatial coherence, for instance using a long multimode fiber and a broadband laser [7-10] or by fabricating an array of mutually incoherent lasers [11], or tailoring the spatial coherence of random lasers [12,13], these techniques cannot be applied to reduce the speckle of an arbitrary off-the-shelf laser operating at any frequency and bandwidth. Instead, most speckle reduction techniques generate different speckle patterns sequentially and perform the speckle averaging in time to reduce the effective speckle contrast. The speckle contrast, $C$, is defined as $C= \sigma_I/<I>$ where $\sigma_I$ is the standard deviation of the intensity and $<I>$ is the average intensity. Human observers can detect speckle with contrast >4% [14]. Since the contrast scales as $C= M^{-1/2}$ where $M$ is the number of independent speckle patterns [3], at least 600 uncorrelated speckle patterns should be averaged for most applications. The problem, then, is how to generate uncorrelated speckle patterns quickly and with a minimum loss of input intensity. In addition, many applications will also have constraints regarding the power consumption, cost, lifetime, and size of the speckle averaging device.

Uncorrelated speckle patterns can be produced by light with different polarizations, wavelengths, incident angles, or incident phase fronts. A simple approach to generate many speckle patterns is using a moving diffuser such as ground glass [15]; however, mechanical systems are typically slow and cumbersome. More recently, the same conceptual approach has been realized using MEMS based deformable mirrors [16], spatial light modulators [17,18], liquid crystals [19], or diffractive optical elements [14,20]. While these methods offer faster speckle contrast reduction, they still require external power and can be expensive. A solution of colloidal particles acts as a moving diffuser without external power. Instead, room temperature Brownian motion is sufficient to quickly generate uncorrelated speckle patterns [2,21]. However, the transmission is low, as most light is scattered away from the forward direction.

In this paper, we developed a technique to overcome the low transmission which has limited the adoption of colloidal solutions in speckle reduction systems. We demonstrated that the transmission of a laser beam through a colloidal dispersion can be enhanced by orders of magnitude. This technique is applicable to any laser, regardless of its frequency, bandwidth, and polarization. It makes a colloidal solution an ideal system to provide low-loss, high-speed speckle contrast reduction, with the advantages of being compact, low cost, robust, and having no power consumption. Further, by controlling the chemistry and electrostatic charge of the particles, colloidal solutions can be engineered to be stable for years enabling a long lifetime [22].

In a solution of colloidal particles, the correlation time of an individual speckle is given as [23]: $t_d = \tau_0 [l_t/L]^2$, where $\tau_0$ is the diffusion time, $l_t$ is the transport mean free path, and $L$ is the system size. This expression holds in the diffusive regime where $L$ is much larger than $l_t$. The diffusion time describes the time required for a particle to diffuse a distance equal to the optical wavelength, $\lambda$, and is defined as $\tau_0 = 1/[k^2 D]$, where $k = 2\pi/\lambda$ is the wavenumber and $D$ is the diffusion coefficient [23]. In this work, we used a colloidal solution consisting of $TiO_2$ particles (Dupont Ti-Pure R-900, radius=205 nm) in water, and employed a continuous-wave (CW) narrow-band laser operating at a wavelength of 532 nm (Coherent Verdi 5 W). Under these conditions, the diffusion time is ~2.9 ms. The density of $TiO_2$ particles in suspension is $2\times10^{10}$ cm$^{-3}$, giving a transport mean free path of ~100 μm. For $L$=5 mm, the decorrelation time is ~1 μs. Recall that to avoid human observation of speckle, the contrast should be reduced to below 4%, requiring ~600 uncorrelated speckle patterns to be averaged. If the incident light is unpolarized, or the sample is depolarizing (which occurs if light undergoes multiple scattering), then the speckle pattern can be decomposed into two orthogonal polarizations and the contrast observed within the decorrelation time will be reduced by a factor of $\sqrt{2}$ [3]. In our system, multiple scattering through the colloidal solution also depolarizes the laser. Since we effectively generate two uncorrelated speckle patterns within the decorrelation time, the integration time required to achieve ~4% contrast is ~300 μs. Of course, if a human observer is directly viewing the image, then we would only need to generate ~600 speckle patterns within the integration time of the human eye, which is typically reported as ~20 ms [21]. However, in the more general case when a camera is used to record fast temporal dynamics, colloidal particles enable speckle reduction at much shorter integration times. Moreover, speckle contrast could be reduced even more quickly by increasing $L$. Unfortunately, increasing $L$ also has the adverse effect of decreasing the transmission, which scales as $l_t/L$ [24]. Thus in our system with $L$ = 5 mm, we expect a transmission of ~0.02. Next, we present a design that greatly improves the transmission which has limited the adoption of colloidal solutions in speckle reduction systems.

## 2. Enhanced transmission through a colloidal solution

As the thickness of the scattering solution increases, more of the light is reflected than transmitted. This can be seen qualitatively in Fig. 1(a), where a laser beam is incident from above a colloidal solution contained in a glass cuvette and the majority of the light is scattered back out the top of the cuvette. To minimize the reflection, we used an optical fiber to deliver the laser light deep into the colloidal solution. We used a single mode fiber in this work, although we expect similar performance from a multimode fiber. In Fig. 1(b), we show the same cuvette with the fiber delivering the laser light into the solution. In this case, more of the light is transmitted than reflected back through the top (since there is more scattering solution above the fiber tip than below it); however, significant amounts of light are lost to the sides. To overcome this issue, we first thought to coat the sidewalls of the cuvette with a metal film to confine light in the scattering solution and keep it from leaking out the sides. However, metals are lossy and the diffusive light could undergo many reflections from the sidewalls and suffer significant absorption. Instead, we opted to coat the sides of the cuvette with a thick layer of white paint which is highly reflective and has negligible absorption. Figure 1(c) shows the optimized speckle reduction system in which a fiber delivers the laser emission into a painted cuvette. The diffuse laser light appears to be efficiently channeled through the clear base of the cuvette.

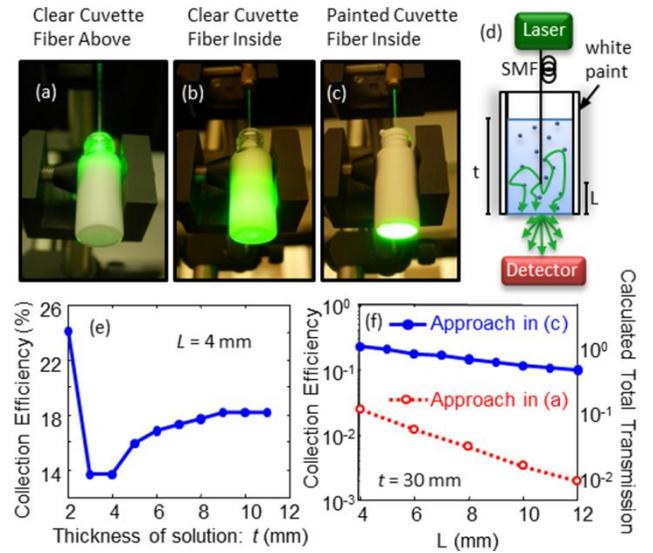

Fig. 1. (Color Online) (a) Photograph of a colloidal solution in a clear glass cuvette. A CW laser at 532 nm is coupled to a single mode fiber and the output from the fiber tip is incident from free space above the colloidal solution of $TiO_2$ particles in water (the particle density is $2x10^{10}$ cm$^{-3}$). The incident light is primarily scattered at the top of the solution with very little propagating through the base of the cuvette. (b) The same colloidal solution in a glass cuvette with the fiber tip delivering the laser light deep inside the colloidal solution. In this case, emission is seen leaving the base of the cuvette, as well as the sides. (c) The same colloidal solution is contained in a cuvette with white paint covering the sides. Again a fiber is used to deliver the laser light deep into the colloidal solution. Strong output is observed through the base of the cuvette. (d) Schematic of the setup shown in (c). The fiber tip is at a distance $L$ from the base of the cuvette and the total thickness of the solution in the cuvette is $t$. After multiple scattering in the colloidal solution, most of the diffusive light escapes through the base of the cuvette, and the output has an angular distribution similar to a Lambertian source. (e) Fraction of light collected by the detector shown in (d) as the colloidal solution was gradually added to the cuvette. The fiber tip was fixed 4 mm from the base of the cuvette. Initially, the transmission decreases as colloidal solution is added until the tip of the fiber is submerged. Adding solution above the fiber tip increases the transmission until it saturates at ~0.18. (f) Measured transmission in (c) was compared to that in (a). The fraction of light collected by the detector (collection efficiency) is denoted on the left axis and the calculated transmission assuming the transmitted light follows a Lambertian distribution is shown on the right axis. The solid blue circles represent the transmission as the fiber was moved away from the cuvette base inside the solution. In this experiment, ~30 mm of solution was placed in the cuvette to ensure the transmission was saturated as observed in (e). The open red circles represent the transmission through different thicknesses of the same colloidal solution contained in an un-painted cuvette with the laser beam incident from free space above the solution.

We then set out to quantitatively evaluate the transmission properties of the optimized system. We first evaluated the thickness of solution required above the fiber to maximize the transmission. This experiment is shown schematically in Fig. 1(d). The fiber was held at a fixed distance from the base of the cuvette ($L$ ~ 4 mm) and we gradually increased the thickness of the solution ($t$). As shown in Fig. 1(e), the intensity measured at the detector originally decreases as solution is added, because more light is scattered back. Once the fiber tip is submerged in the solution, the reflection from the tip of the fiber is reduced due to the lower index contrast between the fiber core and the colloidal solution as compared to the index contrast between the fiber core and air. Adding more solution above the fiber tip further increases the transmission, because light that is initially scattered backwards may be further scattered into the forward direction and escape through the base of the cuvette. The collection efficiency saturates when the thickness of the solution above the fiber ($t-L$) is much larger than the distance to the base of the cuvette ($L$) so that the escape of light through the top surface of the colloidal solution is eliminated.

We then compared the transmission of our optimized system (painted cuvette with emission delivered via fiber) with the transmission through the same colloidal solution in an unpainted glass cuvette with emission incident from free space. In Fig. 1(f), we present the transmission for the two configurations. In the case of the painted cuvette, 30 mm of solution was kept in the cuvette at all times and $L$ describes the distance from the fiber tip to the base of the cuvette. In the case of the clear cuvette where the laser beam is incident from free space, $L$ describes the total thickness of the solution. Using the fiber coupling technique, we observed more than an order of magnitude enhancement of transmission and a slower decay with $L$.

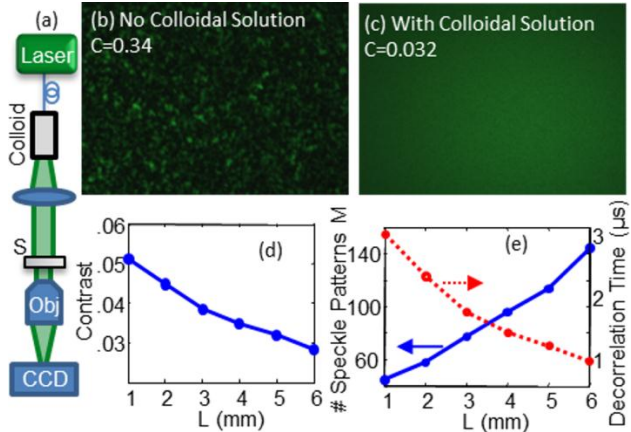

Fig. 2. (Color Online) (a) Schematic of the experimental setup used to measure speckle contrast. The diffused laser emission was collimated by a lens and incident onto a scattering film (S). The transmitted light was imaged onto a CCD camera by an objective lens (Obj). (b) Without a colloidal solution, speckle is observed with contrast $C$= 0.34. (c) Using the fiber-coupled colloidal solution ($L$ = 5mm in a painted cuvette), speckle is suppressed with $C$= 0.032. (d) Measured speckle contrast as a function of the distance $L$ between the fiber tip and the base of the cuvette. The integration time was fixed at 129 µs. (e) Based on the speckle contrast in (d), the number of independent speckle patterns averaged during the 129 µs integration time and the decorrelation time of a speckle pattern were estimated at different $L$.

Our measurement shown in Fig. 1(e) reveals that the fraction of the original laser intensity collected by our detector saturated at ~0.18; however, we did not observe any light penetrating through the paint or reflected back through the top of the cuvette. To account for the rest of the input light, we noticed the finite collection angle of the photodetector, which was placed 1 cm away from the base of cuvette. Since we did not use any collection optics, e.g. a lens, light that leaves the base of the cuvette at shallow angles would miss the detector. We then calculated the fraction of light that reaches the detector assuming the output from any point on the base of the cuvette followed a Lambertian distribution. Based on the 0.8 cm diameter of the cuvette and the 1 cm diameter of the photodetector, we found that only ~20% of the output from the cuvette base would hit the detector. Thus, the combination of painting the cuvette and fiber coupling effectively channels most of the input light through the base of the cuvette. The collection efficiency can be further increased by reducing the cuvette diameter or using collection optics for the diffused output. We expect most of the diffused laser emission could be directed onto a target by employing the efficient means that have been developed for collecting emission from a Lambertian source, e.g., LEDs [25,26,26].

## 3. Speckle contrast reduction using a colloidal solution

Having mitigated the problem of low transmission, we then confirmed that the colloidal solution provided fast speckle contrast reduction. To do this, the output from the base of the cuvette was collimated onto a thin scattering film consisting of dry $TiO_2$ particles on a coverslip using a lens (diameter = 2.54 cm, focal length = 2.54 cm). The far-field pattern of transmitted light was projected onto a charge coupled device (CCD) camera (Moticam 2300 ) by an objective lens (5×, NA=0.1). A schematic of the experimental setup is shown in Fig. 2(a). In the absence of the colloidal solution (i.e. with an empty cuvette) speckle was observed, as shown in Fig. 2(b). The speckle contrast in the absence of the colloidal solution was found to be 0.34. Note that fully developed speckle of single polarization exhibits contrast of 1. The observed contrast was reduced by the combination of several effects. First, the laser emission transmitted through the single mode fiber was unpolarized and thus produced two uncorrelated speckle patterns which reduced the contrast by a factor of $\sqrt{2}$. Second, cracks in the $TiO_2$ film allowed ballistic transport of light which manifests as a constant background in the speckle image, reducing the observed contrast. Finally, the finite pixel size of the camera introduced spatial averaging which also contributed to reduction of the observed speckle contrast. Nonetheless, speckle contrast of 0.34 is clearly prohibitive in imaging applications.

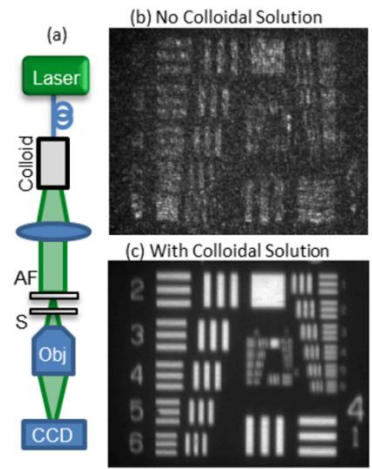

Fig. 3. (Color Online) (a) The CW laser output from the fiber-coupled colloidal solution in a painted cuvette was used to illuminate an AF chart, which was imaged in transmission through a static scattering film (S) onto a CCD camera. (b) Image of the AF chart without the colloidal solution. Speckle formation corrupted the image. (c) Image of the AF chart with the colloidal solution. The speckle contrast was greatly reduced and the features of the AF chart are clearly visible despite the scattering film. The integration time of the CCD camera was adjusted in (b) and (c) to fully utilize the dynamic range of the CCD camera.

We then added 30mm of solution to the cuvette and recorded speckle images with the fiber at varying distances from the base of the cuvette ($L$). The integration time was fixed at 129 µs. Even at this short integration time, the colloidal solution effectively suppressed the speckle, as shown in Fig. 2(c) for $L = 5$ mm. In Fig. 2(d), we plot the speckle contrast, which decreases with $L$. The contrast was calculated as $C = \sigma_I /<I>$ where $\sigma_I$ is the standard deviation of the intensity at each pixel and $<I>$ is the average pixel intensity of the entire image. Based on the measured contrast, we estimated the number of uncorrelated speckle patterns which were averaged as $M = (C_0/C)^2$, where $C_0$ is the speckle contrast recorded without the colloidal solution ($C_0 = 0.34$). Given the number of uncorrelated speckle patterns, we then estimated the time required to generate a new speckle pattern by dividing the integration time by $M$, as shown in Fig. 2(e). We found that the colloidal solution generated uncorrelated speckle patterns in ~1 µs.

Finally, we demonstrated that the ability of the colloidal solution to reduce speckle results in a dramatic improvement of imaging quality. The imaging experiments were performed in transmission, as shown schematically in Fig. 3(a). The emission from the base of the cuvette was directed onto a US Air Force (AF) resolution test chart and subsequently imaged onto a CCD camera. A static scattering film was placed between the AF chart and the imaging objective to simulate imaging through a scattering medium. Without a colloidal solution, speckle is formed throughout the image, as seen in Fig. 3(b), drastically degrading the image. By introducing a colloidal solution, speckle is greatly suppressed and the features on the AF chart can be clearly seen through the static scattering film, which merely increases the background level throughout the image.

## 4. Conclusion

In summary, we have presented a simple and robust technique to achieve high transmission through a colloidal solution enabling fast speckle contrast reduction. The colloidal solution is low cost, compact, requires no external power, and can have a long lifetime. The alignment of the fiber is not critical, allowing for mechanically robust operation, and the device performance is not affected by typical fluctuations in the ambient temperature. The speckle contrast was reduced to below 0.04 at an integration time of 129 µs. We also confirmed that the reduced speckle contrast translated to higher quality images. This work demonstrates the potential for colloidal solutions to be used in a wide range of laser imaging applications.

This work was partly supported by the ACS Petroleum Fund No. S0872-ND6. G. A. was supported by the NSF under Grant No. DBI-1156585 and the Raymond and Beverly Sackler Institute for Biological, Physical, and Engineering Sciences.

## References

1. K. V. Chellappan, E. Erden, and H. Urey, "Laser-based displays: a review," Appl. Opt. **49,** F79-F98 (2010).
2. J. D. Rigden and E. I. Gordon, "The granularity of scattered optical maser light," Proc. Inst. Radio Eng. **50,** 2367-2368 (1962).
3. J. W. Goodman, "Optical Methods for Suppressing Speckle," in *Speckle phenomena in optics*, (Roberts & Co., 2007), pp. 141-186.
4. J. M. Artigas, A. Felipe, and M. J. Buades, "Contrast sensitivity of the visual system in speckle imagery," J. Opt. Soc. Am. A **11,** 2345-2349 (1994).
5. A. G. Geri and L. A. Williams, "Perceptual assessment of laser-speckle contrast," J. Soc. Info. Display **20,** 22-27 (2012).
6. T. S. McKechnie, "Speckle reduction," in *Topics in Applied Physics*, J. C. Dainty, ed. (Springer-Verlag, 1975), pp. 123-170.
7. J. G. Manni and J. W. Goodman, "Versatile method for achieving 1% speckle contrast in large-venue laser projection displays using a stationary multimode optical fiber," Opt. Express **20,** 11288-11315 (2012).
8. A. H. Dhalla, J. V. Migacz, and J. A. Izatt, "Crosstalk rejection in parallel optical coherence tomography using spatially incoherent illumination with partially coherent sources," Opt. Lett. **35,** 2305-2307 (2010).
9. B. Dingel, S. Kawata, and S. Minami, "Speckle reduction with virtual incoherent laser illumination using a modified fiber array," Optik **94,** 132-136 (1993).
10. B. Dingel and S. Kawata, "Laser-diode microscope with fiber illumination," Opt. Commun. **93,** 27-32 (1992).
11. J. Seurin, G. Xu, B. Guo, A. Miglo, Q. Wang, P. Pradhan, J. D. Wynn, V. Khalfin, W. Zou, C. Ghosh, and R. Van Leeuwen,"Efficient vertical-cavity surface-emitting lasers for infrared illumination applications" in *Vertical-Cavity Surface-Emitting Lasers XV* (SPIE, 2011).
12. B. Redding, M. A. Choma, and H. Cao, "Spatial coherence of random laser emission," Opt. Lett **36,** 3404-3406 (2011).
13. B. Redding, M. A. Choma, and H. Cao, "Speckle-free laser imaging using random laser illumination," Nature Photonics **6,** 355-359 (2012).
14. L. Wang, T. Tschudi, T. Halldorsson, and P. R. Petursson, "Speckle reduction in laser projection systems by diffractive optical elements," Appl. Opt. **37,** 1770-1775 (1998).
15. S. Lowenthal and D. Joyeux, "Speckle removal by a slowly moving diffuser associated with a motionless diffuser," J. Opt. Soc. Am. **61,** 847-851 (1971).
16. M. N. Akram, Z. Tong, G. Ouyang, X. Chen, and V. Kartashov, "Laser speckle reduction due to spatial and angular diversity introduced by fast scanning micromirror," Appl. Opt. **49,** 3297-3304 (2010).
17. J. I. Trisnadi, "Hadamard speckle contrast reduction," Opt. Lett. **29,** 11-13 (2004).
18. M. N. Akram, V. Kartashov, and Z. Tong, "Speckle reduction in line-scan laser projectors using binary phase codes," Opt. Lett. **35,** 444-446 (2010).
19. A. L. Andreev, I. N. Kompanets, M. V. Minchenko, E. P. Pozhidaev, and T. B. Andreeva, "Speckle suppression using a liquid-crystal cell," Quant. Electron. **38,** 1166-1170 (2008).
20. G. Ouyang, Z. Tong, M. N. Akram, K. Wang, V. Kartashov, X. Yan, and X. Chen, "Speckle reduction using a motionless diffractive optical element," Opt. Lett. **35,** 2852-2854 (2010).
21. F. Riechert, G. Bastian, and U. Lemmer, "Laser speckle reduction via colloidal-dispersion-filled projection screens," Appl. Opt. **48,** 3742-3749 (2009).
22. I. D. Morrison and S. Ross, *Colloidal Dispersions: Suspensions, Emulsions, and Foams* (Wiley-Interscience, 2002).
23. D. J. Pine, D. A. Weitz, P. M. Chaikin, and E. Herbolzheimer, "Diffusing-wave spectroscopy," Phys. Rev. Lett. **60,** 1134-1137 (1988).
24. M. C. W. van Rossum and Th. M. Nieuwenhuizen, "Multiple scattering of classical waves: microscopy, mesoscopy, and diffusion," Rev. Mod. Phys. **71,** 313-371 (1999).
25. J. J. Chen and C. T. Lin, "Freeform surface design for a light-emitting diode-based collimating lens," Optical Engineering **49,** 093001 (2010).
26. T. Kari, J. Gadegaard, T. Sndergaard, T. G. Pedersen, and K. Pedersen, "Reliability of point source approximations in compact LED lens designs," Opt. Express **19,** A1190-A1195 (2011).